\documentclass[12pt,longbibliography]{iopart}
\expandafter\let\csname equation*\endcsname\relax
\expandafter\let\csname endequation*\endcsname\relax
\usepackage{amsmath}
\usepackage{graphicx}
\usepackage{dcolumn}
\usepackage{bm}
\usepackage{epsfig}
\usepackage{color}
\usepackage{longtable}
\usepackage[numbers,sort&compress]{natbib} 
\usepackage[utf8]{inputenc}
\usepackage{braket}
\usepackage{tikz}
\usetikzlibrary{snakes}
\usepackage{makecell}
\usepackage[flushleft]{threeparttable}
\usepackage[utf8]{inputenc}
\usetikzlibrary{snakes}
\usepackage{xcolor}

\def\sfm#1{\textcolor{black}{#1}}  
\newcommand{\newblock}{\ }
\begin{document}

\title[Delbr\"uck scattering above the pair production threshold]{Delbr\"uck scattering above the pair production threshold: Going beyond the Born approximation}

\author{J Sommerfeldt$^{1,2}$, V A Yerokhin$^{3}$ and A Surzhykov$^{1,2}$}

\address{$^1$Physikalisch-Technische Bundesanstalt, D–38116 Braunschweig, Germany}
\address{$^2$Technische Universität Braunschweig, D–38106 Braunschweig, Germany}
\address{$^3$Max-Planck-Institut für Kernphysik, D–69117 Heidelberg, Germany}
\ead{j.sommerfeldt@tu-braunschweig.de}

\vspace{10pt}
\begin{indented}
\item[]August 2023
\end{indented}

\begin{abstract}
We present a theoretical method to calculate Delbrück scattering amplitudes for photon energies above the electron-positron pair production threshold. The method is based on the application of the relativistic Dirac-Coulomb Green function \sfm{and describes} the interaction of the virtual $e^+e^-$ pair with the Coulomb field of a target to all orders in the coupling strength parameter $\alpha Z$. To illustrate the application of the \sfm{developed} approach, detailed calculations have been performed for the scattering of 2.754~MeV photons off bare ions with a wide range of nuclear charge numbers. Results of these calculations clearly indicate that the higher-order terms beyond the Born approximation lead to a strong enhancement of the imaginary part of the Delbrück amplitude and have to be taken into account for the analysis and guidance of gamma-ray scattering experiments.
\end{abstract}

%
%
%
%
%

\section{Introduction}
Elastic scattering of photons by atomic targets is a well-established experimental technique commonly used to investigate the structure of atoms, molecules and solid state materials~\cite{KANE198675, ROY19993}. For moderate photon energies between 1 and 10~MeV, there are three \sfm{main channels contributing} to this process: elastic nuclear Compton, Rayleigh and Delbrück scattering. The \sfm{Delbrück channel} is the elastic scattering of photons by the Coulomb field of a nucleus via the creation and annihilation of virtual electron positron pairs. \sfm{This process} is of particular interest because it is one of very few non-linear quantum electrodynamical (QED) processes that can be studied in experiment~\cite{MILSTEIN1994183,SCHUMACHER1999101}. 

The accurate theoretical \sfm{description} of the Delbrück process has been a very challenging task in the past, mainly due to the necessity to \sfm{account for} the coupling between the virtual electron-positron pairs and the Coulomb field of a nucleus. \sfm{Previously}, this coupling was taken into account mainly within the lowest-order Born approximation~\cite{PhysRevD.12.206, BARNOY1977132, PhysRevD.26.908, FALKENBERG19921, PhysRevLett.118.204801}. This approximation is based upon expanding the Delbrück amplitude in powers of $\alpha Z$, where $\alpha$ is the fine structure constant and $Z$ is the nuclear charge, and neglecting all terms beyond the lowest order $\sim(\alpha Z)^2$. The Born approximation works very well for light target atoms since $\alpha Z \ll 1$ in this case. For heavy systems, in contrast, higher-order corrections are not negligible anymore and are known to modify the Delbrück amplitude significantly~\cite{SCHUMACHER1999101, SCHUMACHER1975134,RULLHUSEN1979166,rullhusen_coulomb_1979, PhysRevC.23.1375}. To estimate those corrections, a number of approximate methods have been developed in the past, which \sfm{are applicable}, however, only in very restricted parameter regimes. For example, the limit of high photon energies and large or small scattering angles as well as the low photon energy limit have been discussed in the literature~\cite{PhysRev.182.1873, PhysRevD.2.2444, PhysRevD.5.3077,MILSHTEIN1983135,Milstein_1988,PhysRevA.77.032118}. To the best of our knowledge, no general approach was successfully applied to treat Delbrück scattering beyond the Born approximation for arbitrary energies and scattering angles.

Within the framework of quantum electrodynamics, all-order calculations of Delbrück scattering can be performed by using the Dirac-Coulomb propagator which accounts for the Coulomb interaction with a target atom exactly. The structure of this propagator is more complicated than that of the \sfm{free Dirac propagator}, thus making calculations computationally very demanding. Up to now, only \sfm{few calculations were performed} for photon energies below the pair production threshold~\cite{PhysRevD.45.2982,scherdin_coulomb_1995,PhysRevA.105.022804}. The above threshold case is even more demanding due to the fact that \sfm{the production of a real electron-positron pair} is possible in this regime. In the present work, we propose an efficient approach for \sfm{calculations of the above-threshold Delbrück scattering, which} allows to take into the account the interaction of the electron-positron pairs with the Coulomb centre to all orders. Very recently, this approach was applied by us to explain a long-standing discrepancy between experiment and theory for the scattering of 2.754~MeV photons off plutonium targets~\cite{PhysRevLett.131}. 

The present manuscript is organized as follows. In section~\ref{TheorBack}, we recall the basic equations of relativistic quantum electrodynamics used to describe Delbrück scattering. In particular, we discuss the Feynman diagram of the scattering process and the corresponding amplitude. The evaluation of this amplitude \sfm{involves} multidimensional integrals, both, over the energies of the virtual electron and positron as well as over the spatial vertex coordinates. We discuss \sfm{the theoretical approach that makes} the computation of these integrals accessible and numerically stable. In particular, we show that the integration over the energies can be performed using a Wick rotated contour and the treatment of the radial integrals can be simplified by using analytical expressions for the Dirac-Coulomb Green function in the asymptotic regime. The details of the practical implementation of our method as well as estimates of uncertainties introduced by the numerical procedures used in the computation are presented in section~\ref{CompDet}. In section~\ref{NumRes}, we apply our method to calculate the Delbrück amplitude for the scattering of 2.754~MeV photons off bare zinc, cerium, lead and plutonium nuclei. The result of the all-order in $\alpha Z$ calculations were compared with the predictions of the lowest order Born approximation in order to investigate the role of the higher-order Coulomb corrections. We found in particular that these Coulomb corrections beyond the Born approach can significantly modify the imaginary part of the scattering amplitude, thus stressing the importance of the all-order treatment of Delbrück scattering. Finally,  we summarize our work in section~\ref{Conc}. Relativistic units $\hbar = m_e = c = 1$ are used throughout this paper, if not stated otherwise.

\section{Theoretical background} \label{TheorBack}

\begin{figure}
\begin{center}
\begin{tikzpicture} [scale = 0.5]
\draw (0,0) circle (3);
\draw (0,0) circle (3.2);

\draw[fill = black] (3.1,0) circle (0.2);
\draw[fill = black] (-3.1,0) circle (0.2);

\draw[fill=black] (0.35,3.1)--(-0.35,2.8)--(-0.35,3.4);
\draw[fill=black] (-0.35,-3.1)--(0.35,-2.8)--( 0.35,-3.4);

\node at (-2.1,-0.) {$\boldsymbol{r}_1$};
\node at (2.1,-0.) {$\boldsymbol{r}_2$};

\node at (-6, 2.) {$\boldsymbol{k}_1, \boldsymbol{\epsilon}_1$};
\node at (+6, 2.) {$\boldsymbol{k}_2, \boldsymbol{\epsilon}_2$};

\draw[snake=snake,segment amplitude=14pt, segment length = 18pt] (3.1,0) -- (8.2,0);
\draw[snake=snake,segment amplitude=14pt, segment length = 18pt] (-3.1,0) -- (-8.2,0);

\end{tikzpicture}\caption{Feynman diagram for Delbrück scattering to all orders in $\alpha Z$ and leading order in $\alpha$.}\label{FeynmanDel}
\end{center}
\end{figure}
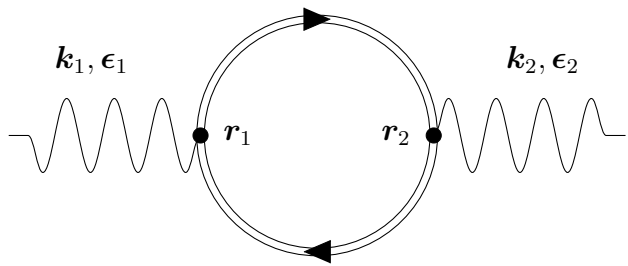

Within the framework of QED, Delbrück scattering can be described by the Feynman diagram in Fig.~\ref{FeynmanDel}. As usual, the wavy lines in the diagram represent the incoming/outgoing photon and the double lines indicate the propagator of the virtual electron and positron moving in the Coulomb field of the nucleus. In this so-called Furry picture, the interactions between the leptons and the nucleus are treated in all-orders in $\alpha Z$. In contrast, the interaction with the photon field is treated in lowest-order in the coupling parameter $\alpha$ displayed by the two vertices $\boldsymbol{r}_1$ and $\boldsymbol{r}_2$.

By using the Feynman correspondence rules, we can translate the diagram in Fig.~\ref{FeynmanDel} into the scattering amplitude
\begin{equation} \label{MatrixElement}
\begin{aligned}
M^{(D)}(\boldsymbol{k}_1,\boldsymbol{\epsilon}_1, \boldsymbol{k}_2,\boldsymbol{\epsilon}_2, Z)  =& \frac{i\alpha}{2\pi} \int_{-\infty}^\infty \text{d}z\int_{-\infty}^\infty \text{d}z'\int \text{d}^3\boldsymbol{r}_1\int \text{d}^3\boldsymbol{r}_2~\text{Tr}\Big[R(\boldsymbol{r}_1,\boldsymbol{k}_1,\boldsymbol{\epsilon}_1) G(\boldsymbol{r}_1,\boldsymbol{r}_2,z)\\
&\times R^\dagger(\boldsymbol{r}_2,\boldsymbol{k}_2,\boldsymbol{\epsilon}_2)G(\boldsymbol{r}_2,\boldsymbol{r}_1,z')\Big]\delta (\omega+z-z')~,\\
\end{aligned}
\end{equation}
where $z$ and $z'$ are the energy arguments of the electron propagators and $\omega$ is the energy of the incoming and outgoing photon. Moreover, two theoretical ``ingredients'' enter into the scattering amplitude: The electron-photon interaction operator $R(\boldsymbol{r},\boldsymbol{k},\boldsymbol{\epsilon})$ and the Dirac-Coulomb Green function $G(\boldsymbol{r}_2,\boldsymbol{r}_1,z)$. In what follows, we will discuss these functions in detail starting with the latter one. For the numerical analysis of amplitude~\eqref{MatrixElement}, it is convenient to expand the Green function into partial waves
\begin{equation} \label{AnalyticalGreen}
\begin{aligned}
\!
G&(\boldsymbol{r}_2,\boldsymbol{r}_1,z) = \sum_{\kappa\mu} \frac{1}{w_\kappa(z)}\\ 
&\times\Bigg[\Theta(r_2-r_1) \left(\begin{array}{c}
F_{\kappa,\infty}^1 (r_2,z) \chi_\kappa^\mu(\hat{\boldsymbol{r}}_2)\\
iF_{\kappa,\infty}^2 (r_2,z) \chi_{-\kappa}^\mu(\hat{\boldsymbol{r}}_2)
\end{array}\right)
\left(\begin{array}{cc}
F_{\kappa,0}^1 (r_1,z) \chi_\kappa^{\mu\dagger}(\hat{\boldsymbol{r}}_1) & -iF_{\kappa,0}^2 (r_1,z) \chi_{-\kappa}^{\mu\dagger}(\hat{\boldsymbol{r}}_1)
\end{array}\right)\\
&+ \Theta(r_1-r_2) \left(\begin{array}{c}
F_{\kappa,0}^1 (r_2,z) \chi_\kappa^\mu(\hat{\boldsymbol{r}}_2)\\
iF_{\kappa,0}^2 (r_2,z) \chi_{-\kappa}^\mu(\hat{\boldsymbol{r}}_2)
\end{array}\right)
\left(\begin{array}{cc}
F_{\kappa,\infty}^1 (r_1,z) \chi_\kappa^{\mu\dagger}(\hat{\boldsymbol{r}}_1) & -iF_{\kappa,\infty}^2 (r_1,z) \chi_{-\kappa}^{\mu\dagger}(\hat{\boldsymbol{r}}_1)
\end{array}\right)\Bigg]\\
\end{aligned}
\end{equation}
which are characterized by \sfm{the Dirac angular-momentum quantum number} $\kappa$ and total angular momentum projection $\mu$. In Eq.~\eqref{AnalyticalGreen}, moreover, $F^{1,2}_{\kappa, 0}$ and $F^{1,2}_{\kappa, \infty}$ are solutions of the radial Dirac equation that are regular at the origin and at infinity, $\chi_\kappa^{\mu\dagger}$ are the spin-angular wave functions and the Wronskian is given by 
\begin{equation} \label{Wronskian}
w_\kappa(z) = r^2 [F^2_{\kappa,0}(r,z) F^1_{\kappa,\infty}(r,z)-F^1_{\kappa,0}(r,z) F^2_{\kappa,\infty}(r,z)]~.
\end{equation} 
For a pure Coulomb potential, a closed analytical form of the radial components of the Green function is known 
\begin{subequations} \label{GreenCoulomb}
\begin{align}
\left[\begin{array}{c}
F^1_{\kappa,0}(x,z)\\
F^2_{\kappa,0}(x,z)
\end{array}\right] &= \left[\begin{array}{c}
\frac{\sqrt{1+z}}{2cx^{3/2}} \left((\lambda-\nu)M_{\nu-1/2,\lambda}(2cx) - \left(\kappa-\frac{\gamma}{c} \right)M_{\nu+1/2,\lambda}(2cx)\right)\\ [0.2cm]
\frac{\sqrt{1-z}}{2cx^{3/2}} \left((\lambda-\nu)M_{\nu-1/2,\lambda}(2cx) + \left(\kappa-\frac{\gamma}{c} \right)M_{\nu+1/2,\lambda}(2cx)\right)\\
\end{array}\right]~,\hphantom{\Gamma(\lambda-\nu)} \label{GreenCoulomba}\\
\left[\begin{array}{c}
F^1_{\kappa,\infty}(x,z)\\
F^2_{\kappa,\infty}(x,z)
\end{array}\right] &= \frac{\Gamma(\lambda-\nu)}{\Gamma(1+2\lambda)}\left[\begin{array}{c}
\frac{\sqrt{1+z}}{2cx^{3/2}} \left(\left(\kappa+\frac{\gamma}{c} \right)W_{\nu-1/2,\lambda}(2cx) + W_{\nu+1/2,\lambda}(2cx)\right)\\ [0.2cm]
\frac{\sqrt{1-z}}{2cx^{3/2}} \left(\left(\kappa+\frac{\gamma}{c} \right)W_{\nu-1/2,\lambda}(2cx) - W_{\nu+1/2,\lambda}(2cx)\right)\\
\end{array}\right]~, \label{GreenCoulombb}
\end{align}
\end{subequations}
where $M_{\kappa,\mu}$ and $W_{\kappa,\mu}$ are the Whittaker functions, $c = \sqrt{1-z^2}$, $\gamma = \alpha Z$, $\nu = \gamma z/c$ and $\lambda = \sqrt{\kappa^2-\gamma^2}$, see Ref.~\cite{MOHR1998227} for further details. Here, the branch of the square root is taken so that Re$(c) \geq 0$ and, moreover, the Wronskian~\eqref{Wronskian} is unity, $w_\kappa(z) = 1$.

As mentioned above, besides the Green function, the electron-photon interaction operator $\hat{R}(\boldsymbol{r},\boldsymbol{k}, \boldsymbol{\epsilon})$ with wave vector $\boldsymbol{k}$ and polarization vector $\boldsymbol{\epsilon}$ also appears in Eq.~\eqref{MatrixElement}. Similar to $G(\boldsymbol{r}_2,\boldsymbol{r}_1,z)$, it is convenient to expand this operator into its multipole components. In Coulomb gauge and in the helicity basis for the photon polarization $\boldsymbol{\epsilon}_\lambda = \tfrac{1}{\sqrt{2}}(\boldsymbol{e}_x + i \lambda \boldsymbol{e}_y)$, $\lambda = \pm 1$, this expansion reads as
\begin{equation} \label{ElPhOp}
\begin{aligned}
\hat{R}(\boldsymbol{r},\boldsymbol{k},\boldsymbol{\epsilon}) = \boldsymbol{\alpha}\cdot\boldsymbol{\epsilon}_\lambda e^{i\boldsymbol{k}\cdot\boldsymbol{r}}= &\sqrt{2\pi}\sum_{PLM} i^L  \sqrt{2L+1} (i\lambda)^P D^L_{M\lambda}(\boldsymbol{\hat{k}}) \boldsymbol{\alpha}\cdot \boldsymbol{a}_{LM}^{(P)}~.
\end{aligned}
\end{equation}
Here, the magnetic ($P = 0$) and electric ($P = 1$) multipole fields are given by
\begin{subequations} \label{MultComp}
\begin{align} 
\boldsymbol{a}^{(0)}_{LM} &=  j_{L}(\omega r)~\boldsymbol{T}_{LLM}~,\\
\begin{split}
\boldsymbol{a}^{(1)}_{LM} &= \sqrt{\frac{L+1}{2L+1}}j_{L-1}(\omega r)~\boldsymbol{T}_{L,L-1,M} - \sqrt{\frac{L}{2L+1}}j_{L+1}(\omega r)~\boldsymbol{T}_{L,L+1,M}~,
\end{split}
\end{align}
\end{subequations}
with $j_L$ being the spherical Bessel function, and the vector spherical harmonics $\boldsymbol{T}_{JLM}$ are constructed as irreducible tensors of rank $J$ as
\nopagebreak
\begin{equation}
\boldsymbol{T}_{JLM} = \sum_{\mu = -1}^1 \langle L~(M-\mu)~1~\mu\vert J~M\rangle Y_{L,M-\mu}\boldsymbol{\xi}_\mu~,
\end{equation}
see Ref.~\cite{rose_elementary_1957} for further details. By inserting Eqs.~\eqref{AnalyticalGreen} and \eqref{ElPhOp} into amplitude~\eqref{MatrixElement} one can solve all angular integrals analytically while the integration over the radial coordinates and the energy needs to be performed numerically~\cite{PhysRevA.105.022804}.  

\subsection{Energy integral}
\begin{figure} 
\begin{center}
\includegraphics[width = 1.\linewidth]{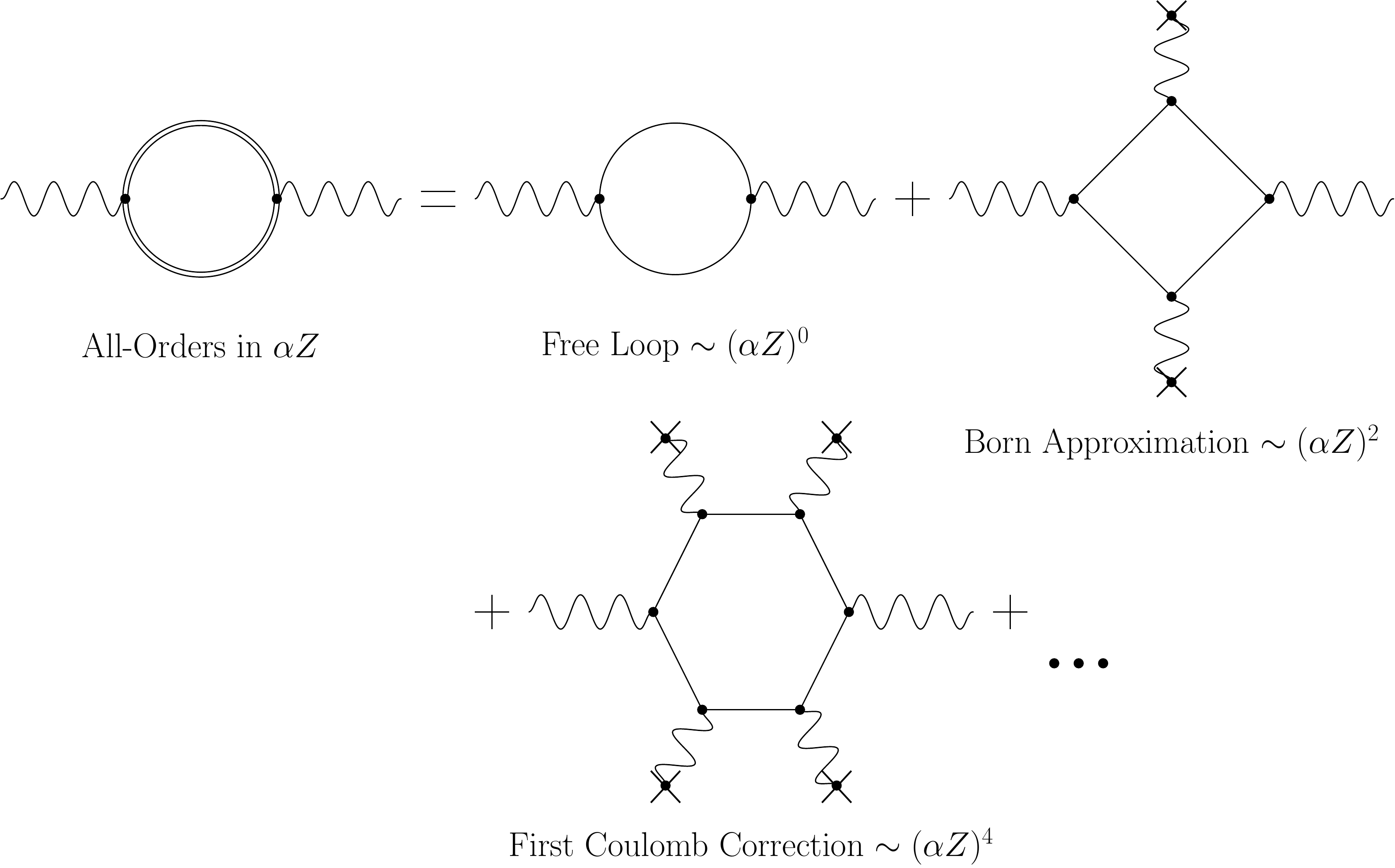} 
\caption{Expansion of the full Delbrück scattering Feynman diagram (left-hand side) into Coulomb interactions (right-hand side). The expansion consists of the free-loop diagram (zeroth-order in $\alpha Z$), the lowest-order Born approximation (second-order in $\alpha Z$) and the Coulomb corrections (fourth-order in $\alpha Z$ and higher). We only show one diagram for each order although all permutations of the photon interactions should also be taken into account.} \label{delbrueck_expansion}
\end{center}
\end{figure}

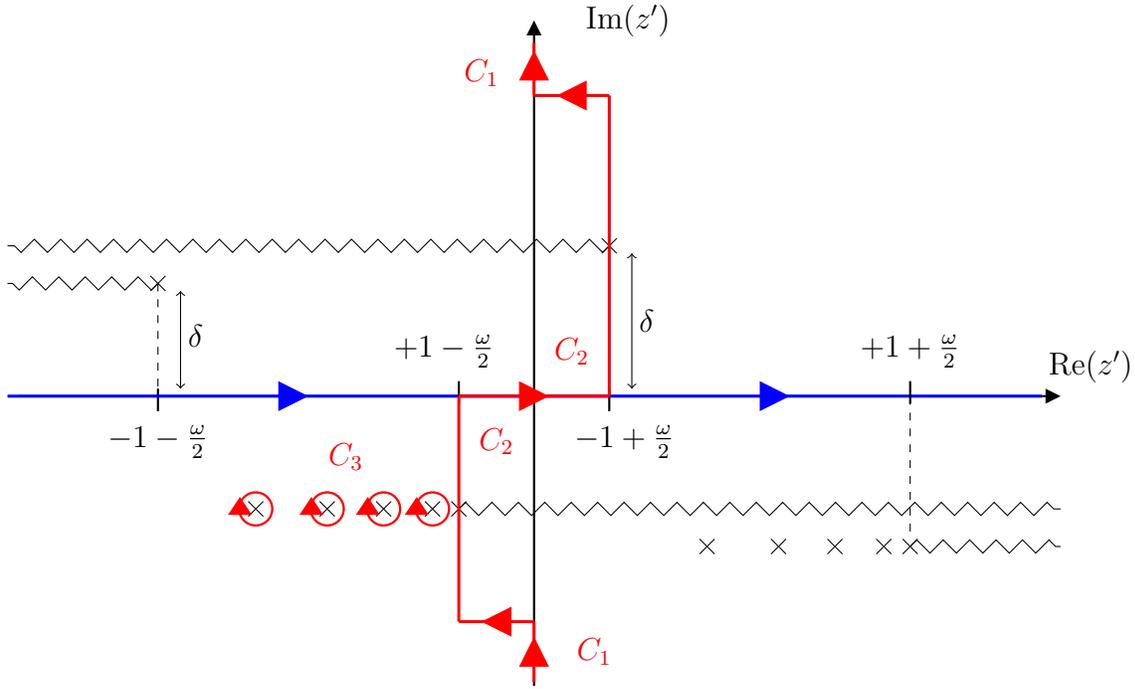
\begin{figure}
\begin{center}
\begin{tikzpicture} [scale = 1.]
\node (org) at (0.0,  0.0) {}; 
\node (xmin) at (-7.0,  0.0) {}; 
\node (xmax) at (7.0, 0.0) {}; 
\node (ymax) at ( 0.0, -4.) {}; 
\node (ymin) at (0.0, 5.) {}; 

\draw[thick] (xmin) -- (xmax);
\draw[thick] (ymin) -- (ymax);
\fill[fill=black] (0.0,5.0)--(-0.1, 4.8)--(0.1, 4.8);
\fill[fill=black] (7.0,0.0)--(6.8, -0.1)--(6.8, 0.1);
\node at (1.25,5.0) {Im$(z')$};
\node at (7.4,0.4) {Re$(z')$};

\draw (-5.1,1.4) -- (-4.9,1.6);
\draw (-5.1,1.6) -- (-4.9,1.4);
\draw[dashed] (-5.0,1.5) -- (-5.0,0.0);
\draw[thick] (-5.0,0.1) -- (-5.0,-0.2);
\node at (-5.0,-0.6) {$-1 - \frac{\omega}{2}$};
\draw[snake=zigzag] (-5.0,1.5) -- (-7.0,1.5);

\draw (0.9,1.9) -- (1.1,2.1);
\draw (0.9,2.1) -- (1.1,1.9);
\draw[dashed] (1.0,1.8) -- (1.0,0.0);
\draw[thick] (1.0,0.1) -- (1.0,-0.2);
\node at (1.2,-0.6) {$-1 + \frac{\omega}{2}$};
\draw[snake=zigzag] (1.0,2.0) -- (-7.0,2.0);

\draw[<->] (-4.7,0.1) -- (-4.7,1.4);
\node at (-4.5,0.8) {$\delta$};

\draw[<->] (1.3,0.1) -- (1.3,1.9);
\node at (1.5,1.0) {$\delta$};

\draw (-3.6,-1.4) -- (-3.8,-1.6);
\draw (-3.6,-1.6) -- (-3.8,-1.4);

\draw[line width = 0.3mm,red] (-3.7,-1.5) circle (0.22);
\fill[fill=red] (-3.75,-1.58)--(-4.07, -1.58)--(-3.91, -1.34);

\draw (-2.65,-1.4) -- (-2.85,-1.6);
\draw (-2.65,-1.6) -- (-2.85,-1.4);

\draw[line width = 0.3mm,red] (-2.75,-1.5) circle (0.22);
\fill[fill=red] (-2.80,-1.58)--(-3.12, -1.58)--(-2.96, -1.34);

\draw (-1.9,-1.4) -- (-2.1,-1.6);
\draw (-1.9,-1.6) -- (-2.1,-1.4);

\draw[line width = 0.3mm,red] (-2.0,-1.5) circle (0.22);
\fill[fill=red] (-2.05,-1.58)--(-2.37, -1.58)--(-2.21, -1.34);

\draw (-1.25,-1.4) -- (-1.45,-1.6);
\draw (-1.25,-1.6) -- (-1.45,-1.4);

\draw[line width = 0.3mm,red] (-1.35,-1.5) circle (0.22);
\fill[fill=red] (-1.40,-1.58)--(-1.72, -1.58)--(-1.56, -1.34);

\draw (2.4,-1.9) -- (2.2,-2.1);
\draw (2.4,-2.1) -- (2.2,-1.9);

\draw (3.35,-1.9) -- (3.15,-2.1);
\draw (3.35,-2.1) -- (3.15,-1.9);

\draw (4.1,-1.9) -- (3.9,-2.1);
\draw (4.1,-2.1) -- (3.9,-1.9);

\draw (4.75,-1.9) -- (4.55,-2.1);
\draw (4.75,-2.1) -- (4.55,-1.9);

\draw[line width = 0.4mm,blue] (-7.0,0.0) -- (6.75,0.0);
\fill[fill=blue] (3.4,0.0)--(3.0, -0.2)--(3.0, 0.2);
\fill[fill=blue] (-3.0,0.0)--(-3.4, -0.2)--(-3.4, 0.2);

\draw (-0.9,-1.4) -- (-1.1,-1.6);
\draw (-0.9,-1.6) -- (-1.1,-1.4);
\draw[dashed] (-1.0,-1.5) -- (-1.0,0.0);
\draw[thick] (-1.0,-0.1) -- (-1.0,+0.2);
\node at (-1.2,+0.6) {$+1-\frac{\omega}{2}$};
\draw[snake=zigzag] (-1.0,-1.5) -- (+7.0,-1.5);

\draw (5.1,-1.9) -- (4.9,-2.1);
\draw (5.1,-2.1) -- (4.9,-1.9);
\draw[dashed] (5.0,-1.8) -- (5.0,0.0);
\draw[thick] (5.0,-0.1) -- (5.0,+0.2);
\node at (5.0,+0.6) {$+1+\frac{\omega}{2}$};
\draw[snake=zigzag] (5.0,-2.0) -- (+7.0,-2.0);

\draw[line width = 0.4mm,red] (0.0,4.7) -- (0.0,4.);
\draw[line width = 0.4mm,red] (0.0,4.)  -- (1.0,4.);
\draw[line width = 0.4mm,red] (1.0,0.0)  -- (1.0,4.);
\draw[line width = 0.4mm,red] (1.0,0.0) -- (-1.0,0.0);
\draw[line width = 0.4mm,red] (-1.0,0.0) -- (-1.0,-3.0);
\draw[line width = 0.4mm,red] (-1.0,-3.0) -- (0.0,-3.0);
\draw[line width = 0.4mm,red] (0.0,-3.0) -- (0.0,-3.8);

\fill[fill=red] (0.2,0.0)--(-0.2, -0.2)--(-0.2, 0.2);
\fill[fill=red] (-0.7,-3.0)--(-0.3, -2.8)--(-0.3,-3.2);
\fill[fill=red] (0.3,4.0)--(0.7, 3.8)--(0.7,4.2);

\fill[fill=red] (0.0,4.6)--(-0.2, 4.2)--(0.2,4.2);
\fill[fill=red] (0.0,-3.2)--(-0.2, -3.6)--(0.2,-3.6);

\node at (0.8,-3.4) {\textcolor{red}{$C_1$}};
\node at (-0.7,4.3) {\textcolor{red}{$C_1$}};

\node at (0.5,0.6) {\textcolor{red}{$C_2$}};
\node at (-0.5,-0.6) {\textcolor{red}{$C_2$}};

\node at (-2.5,-0.8) {\textcolor{red}{$C_3$}};

\end{tikzpicture}
\end{center}\caption{Original (blue) and Wick rotated (red) path for the $z'$ integration in Eq.~\eqref{MatrixElement}. The singularities (black crosses) and branch cuts (zig zag lines) of the integrand are shown in the complex $z'$ plane for $\omega > 2$.}\label{IntPath}
\end{figure}

In order to perform an integration over the energy arguments $z$ and $z'$ of the electron propagators in the amplitude~\eqref{MatrixElement}, it is convenient to consider first a formal $\alpha Z$ expansion of $M^{(D)}$. This expansion is illustrated in terms of Feynman diagrams in Fig.~\ref{delbrueck_expansion}, where the single solid line represents the electron propagator in the absence of an external field while the wavy lines originating from crossed vertices describe single interactions with the Coulomb centre. The first term on the right-hand side of the figure is known as the free-loop contribution which contains a logarithmically divergent loop-momentum integral.
Since this free-loop diagram does not contribute to the light-light interaction process, it should be subtracted from the further calculations as discussed in Ref.~\citep{PhysRevA.105.022804}. In practise, this subtraction can be performed as
\begin{equation}~\label{delbrueck_amplitude_finite}
\widetilde{M}^{(D)}(\boldsymbol{k}_1,\boldsymbol{\epsilon}_1, \boldsymbol{k}_2,\boldsymbol{\epsilon}_2, Z) = M^{(D)}(\boldsymbol{k}_1,\boldsymbol{\epsilon}_1, \boldsymbol{k}_2,\boldsymbol{\epsilon}_2, Z) - M^{(D)}(\boldsymbol{k}_1,\boldsymbol{\epsilon}_1, \boldsymbol{k}_2,\boldsymbol{\epsilon}_2, Z=0)~,
\end{equation}
where $M^{(D)}(\boldsymbol{k}_1,\boldsymbol{\epsilon}_1, \boldsymbol{k}_2,\boldsymbol{\epsilon}_2, Z=0)$ is the amplitude for the free-loop diagram. The remaining amplitude $\widetilde{M}^{(D)}$ accounts for all interactions of the virtual electron-positron pair starting from the lowest order $(\alpha Z)^2$ and is known to be finite. For this amplitude, the integration over $z$ can be carried out trivially due to the presence of the Dirac delta function $\delta (\omega + z - z')$ while the integration over $z'$ is performed numerically. We improve the stability of this integration by substituting $z' \to z'+\frac{\omega}{2}$ thus making the integrand symmetric with respect to the origin $z' = 0$. The resulting integrand is analytical in the entire complex plane except for two sets of branch cuts starting at $z' = \pm 1 + \tfrac{\omega}{2}$ and $z' = \pm 1 - \tfrac{\omega}{2}$ that originate from the positive and negative energy continuum states of the two virtual particles. Moreover, the bound electron states result in two sets of poles at $z' = (\lambda' + n)/\sqrt{\gamma^2 + (\lambda'+n)^2} + \frac{\omega}{2}$ and $z' = (\lambda + n)/\sqrt{\gamma^2 + (\lambda+n)^2} - \frac{\omega}{2}$ with $n = 0,1,2,...$, see Fig.~\ref{IntPath}.
The $z'$ integration on the interval $(-\infty, \infty)$ displayed by the blue path in Fig.~\ref{IntPath} is troublesome since the poles and branch cuts are located infinitely close to the real axis, $\delta \to 0$. To overcome this problem, we construct a new integration path by forming a closed contour in the complex $z'$-plane \sfm{using three parts $C_1$, $C_2$ and $C_3$, see Fig.~\ref{IntPath}. The part $C_1$ extends along the imaginary axis, $C_2$ goes along the cuts of the electron propagators and $C_3$ encloses all bound-state poles of one of the propagators.} By using Cauchy's integration formula, we arrive at the replacement
\begin{equation} \label{intReplace}
\widetilde{M}^{(D)}(\boldsymbol{k}_1,\boldsymbol{\epsilon}_1, \boldsymbol{k}_2,\boldsymbol{\epsilon}_2, Z) = \int_{-\infty}^{\infty} \text{d}z' f(z') = \int_{C_1,C_2} \text{d}z' f(z') - 2\pi i \sum_{z'_n \in C_3} \text{Res}[f,z'_n]~,
\end{equation}
where we used the shorthand notation
\begin{equation} \label{integrandf}
\begin{split}
f(z') = \frac{i\alpha}{2\pi} \int\!\text{d}^3&\boldsymbol{r}_1\!\int\!\text{d}^3\boldsymbol{r}_2~\text{Tr}\Big[R(\boldsymbol{r}_1,\boldsymbol{k}_1,\boldsymbol{\epsilon}_1) G(\boldsymbol{r}_1,\boldsymbol{r}_2,z'-\tfrac{\omega}{2})R^\dagger(\boldsymbol{r}_2,\boldsymbol{k}_2,\boldsymbol{\epsilon}_2)G(\boldsymbol{r}_2,\boldsymbol{r}_1,z'+\tfrac{\omega}{2})\\
&-\left.\Big\{R(\boldsymbol{r}_1,\boldsymbol{k}_1,\boldsymbol{\epsilon}_1) G(\boldsymbol{r}_1,\boldsymbol{r}_2,z'-\tfrac{\omega}{2})R^\dagger(\boldsymbol{r}_2,\boldsymbol{k}_2,\boldsymbol{\epsilon}_2)G(\boldsymbol{r}_2,\boldsymbol{r}_1,z'+\tfrac{\omega}{2})\Big\}\right\vert_{Z=0}\Big]
\end{split}
\end{equation}
for the integrand in amplitude~\eqref{delbrueck_amplitude_finite} and, moreover, $\text{Res}[f,z'_n]$ is the residue of $f(z')$ at its $n$th enclosed pole $z'_n$. While the techniques used to calculate the integral along the imaginary axis $C_1$ are identical to those used in the below-threshold calculations in Ref.~\cite{PhysRevA.105.022804}, the contributions $C_2$ and $C_3$ require some more attention. To calculate the residue $C_3$, we note that the enclosed poles originate from the prefactor $\Gamma(\lambda'-\nu')$ arising in the radial components of the Green function in Eq.~\eqref{GreenCoulombb}. Therefore, to obtain the contribution from the bound states, we simply replace this prefactor by its residue
\nopagebreak
\begin{equation}
\begin{aligned}
\text{Res}\left[\Gamma(\lambda' - \nu'), z'_n = \frac{\lambda' + n}{\sqrt{\gamma^2 + (\lambda' + n)^2}} - \frac{\omega}{2}\right]\\ 
= -\frac{(-1)^n}{n!} \frac{\left(1- \frac{(\lambda' + n)^2}{\gamma^2 + (\lambda' + n)^2}\right)^{3/2}}{\gamma}~.
\end{aligned}
\end{equation}

\noindent The main difficulty in performing the numerical integration along the paths $C_2$ comes from the fact that the integrand is sharply peaked at the beginning of the branch cuts at $z' = \pm 1 \mp \tfrac{\omega}{2}$. To obtain $\int_{C_2} \text{d}z' f(z')$ in a numerically stable way, we use Gauss-Legendre quadrature with an enhanced density of integration points close to the peaked regions by making the substitution $z' \to \pm u^2 + (\pm 1 \mp \tfrac{\omega}{2})$.

\subsection{Radial integrals}
To obtain the integrand $f(z')$ in Eq.~\eqref{intReplace}, one has to perform first the integration over the vertex coordinates $\boldsymbol{r}_1$ and $\boldsymbol{r}_2$, see Eq.~\eqref{integrandf}. As mentioned already above, this six-dimensional integral can be reduced to a two-dimensional one by using the multipole expansions of $R(\boldsymbol{r},\boldsymbol{k},\boldsymbol{\epsilon})$ and $G(\boldsymbol{r}_2,\boldsymbol{r}_1,z)$ and by solving the angular integrals analytically. As shown in Appendix A of Ref.~\citep{PhysRevA.105.022804}, the remaining radial integrals can be traced back to the expression
\begin{equation} \label{IntI}
\begin{aligned}
\mathcal{J} = \int_0^\infty \frac{\text{d}r_1}{r_1} &W_{\nu'+\frac{p_1}{2},\lambda'}(2c'r_1) j_{L_1}(\omega r_1) W_{\nu+\frac{p_2}{2},\lambda}(2cr_1)\\
&\times\int_0^{r_1} \frac{\text{d}r_2}{r_2}M_{\nu'+\frac{p_3}{2},\lambda'}(2c'r_2) j_{L_2}(\omega r_2) M_{\nu+\frac{p_4}{2},\lambda}(2cr_2)~,\\
\end{aligned}
\end{equation}
where $L_1$ and $L_2$ denote the \sfm{multipolarities} of the incoming and outgoing photon, $\lambda$, $\lambda'$ and $\nu$, $\nu'$ are the parameters from Eq.~\eqref{GreenCoulomb} of the two electron propagators and $p_1,...,p_4$ can take the values $+1$ or $-1$. 

To evaluate Eq.~\eqref{IntI}, the integration over $r_1$ and $r_2$ is done numerically for the interval $0 \leq r_1 \leq R_1$, $0 \leq r_2 \leq R_2$, while an analytical integration is carried out in the asymptotic regime, $r_1 > R_1$, $r_2 > R_2$, by using the expansions
\begin{subequations} \label{asympWhit}
\begin{align}
\begin{split}
M_{\alpha,\beta}(2\tilde{c}r) \sim& \frac{\Gamma(1+2\beta)}{\Gamma(\frac{1}{2}+\beta-\alpha)} e^{\tilde{c}r}(2\tilde{c}r)^{-\alpha}\sum_{s=0}^\infty u_M(s,\alpha,\beta)(2\tilde{c}r)^{-s}\\
&+ \frac{\Gamma(1+2\beta)}{\Gamma(\frac{1}{2}+\beta+\alpha)} e^{-\tilde{c}rz+q(\frac{1}{2}+\beta-\alpha)\pi i}(2\tilde{c}r)^{\alpha}\sum_{s=0}^\infty \widetilde{u}_M(s,\alpha,\beta)(-2\tilde{c}r)^{-s}~,\label{asympWhitA}\\
\end{split}\\
&W_{\alpha,\beta}(2\tilde{c}r) \sim e^{-\tilde{c}r}(2\tilde{c}r)^{\alpha}\sum_{s=0}^\infty u_W(s,\alpha,\beta)(-2\tilde{c}r)^{-s}~,
\end{align}
\end{subequations}
where
\begin{subequations}
\begin{align}
u_M(s,\alpha,\beta) &= \frac{(\frac{1}{2}-\beta+\alpha)_s(\frac{1}{2}+\beta+\alpha)_s}{s!}~,\\
\widetilde{u}_M(s,\alpha,\beta) &= \frac{(\frac{1}{2}+\beta-\alpha)_s(\frac{1}{2}-\beta-\alpha)_s}{s!}~,\\
u_W(s,\alpha,\beta) &= \frac{(\frac{1}{2}+\beta-\alpha)_s(\frac{1}{2}-\beta-\alpha)_s}{s!}~,
\end{align}
\end{subequations}
with $q$ being $-1$ for Im$(z) < 0$ and $+1$ otherwise, and $(x)_n$ being the Pochhammer symbol. In the asymptotic regime, moreover, we use the exact expansion of the spherical Bessel function
\begin{equation} \label{asympBes}
\begin{aligned}
j_L(x) &= \sum_{m=0}^L \frac{(L+m)!}{m!(L-m)!}i^{L+1-m}(2x)^{-m-1}[(-1)^{L+1-m}e^{ix}+e^{-ix}]~.
\end{aligned}
\end{equation}
By inserting Eqs.~\eqref{asympWhit} - \eqref{asympBes} into Eq.~\eqref{IntI}, we can express the integral $\mathcal{J}$ as a sum of weighted incomplete gamma functions, see Appendix A for further details. 

While the approach based on the asymptotic expansions~\eqref{asympWhit} - \eqref{asympBes} can be generally performed for incident photon energies both, below and above the electron-positron pair production threshold, its practical realization for the latter case is more cumbersome. To understand this difference, one has to inspect the expansion~\eqref{asympWhitA} of the Whittaker function. As seen from this formula, $M_{\alpha,\beta}$ has an argument $2\tilde{c}r$, where $\tilde{c} = \sqrt{1-(z'\pm\tfrac{\omega}{2})^2}$, which always has a large real part for $\omega < 2$. Therefore, for the below threshold case, the second term in Eq.~\eqref{asympWhitA} is exponentially suppressed compared to the first term and can be neglected for all practical purposes. For $\omega \geq 2$, in contrast, the argument $2\tilde{c}r$ can be purely imaginary leading to the fact that the two terms on the right-hand side of Eq.~\eqref{asympWhitA} are similar in size and both of them need to be taken into account in the integration. This causes a longer computation time compared to the low-energy case.

\section{Computational Details} \label{CompDet}

\begin{figure}
\begin{center}
\includegraphics[width=1.\textwidth]{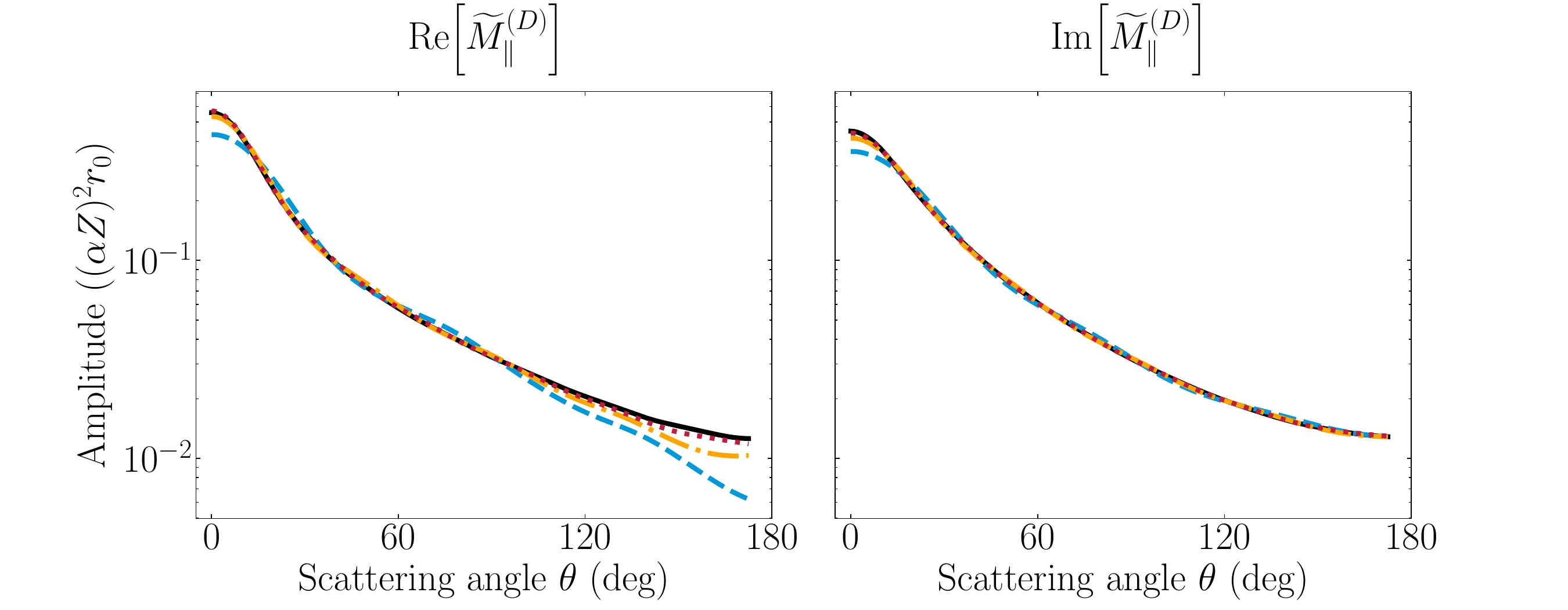} \caption{\sfm{Real (left panel) and imaginary (right panel) part of the Delbrück amplitude~\eqref{delbrueck_amplitude_finite} for the scattering of 2.754~MeV photons by bare plutonium nuclei. The calculations were done for radiation that is linearly polarized within the scattering plane and by summing over the Dirac quantuam number $\kappa$ up to $\kappa = \pm 6$ (blue dashed line), $\kappa = \pm 9$ (orange dash-dotted line), $\kappa = \pm 12$ (red dotted line), and $\kappa = \pm 15$ (black solid line). Results are given in units $(\alpha Z)^2r_0$, where $r_0 = 2.818$ fm is the classical electron radius.}} \label{ConvK}
\end{center}
\end{figure}

In the previous section, we have discussed the evaluation of the Delbrück scattering amplitude for photon energies above the pair production threshold. The practical implementation of our theoretical approach also comes with some difficulties that need to be addressed. The first difficulty is due to the spurious contributions to the real part of the Delbrück amplitude $\widetilde{M}^{(D)}$ which arise from the integration of the \textit{individual} segments $C_1$, $C_2$ and $C_3$ of the energy contour in Fig.~\ref{IntPath}.
\begin{figure}
\begin{center}
\includegraphics[width=0.95\textwidth]{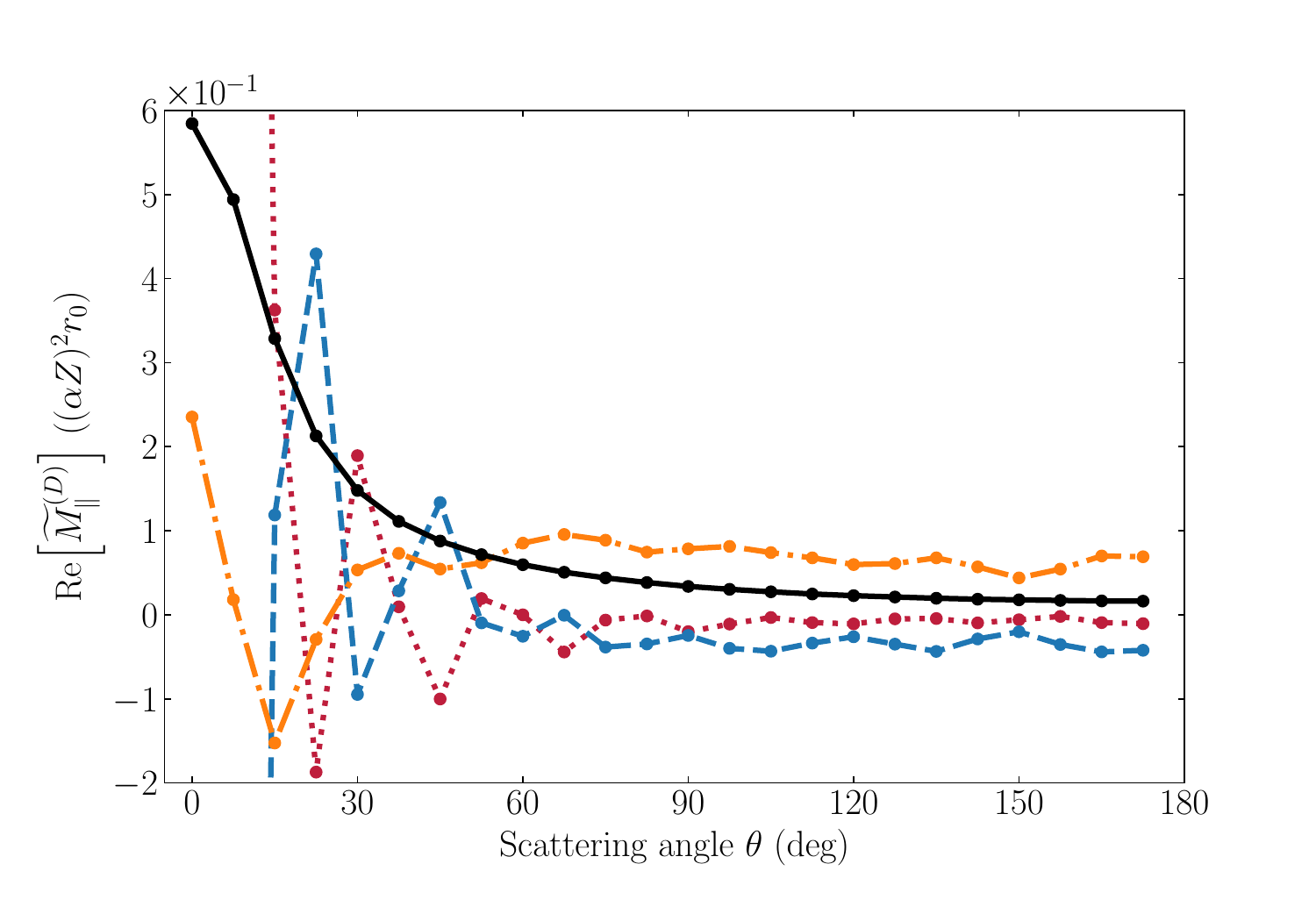} \caption{Real part of the contributions to the Delbrück amplitude~\eqref{intReplace} from the integrals along $C_1$ (orange dash-dotted line), along $C_2$ (blue dashed line), the contributions from the residue (red dotted line) and their sum (black solid line) when summing $\kappa$ in the range from $-15$ to $+15$. \sfm{The amplitudes where obtained for the scattering of 2.754~MeV photons off bare plutonium ions and for radiation that is linearly polarized within the scattering plane spanned by $\boldsymbol{k}_1$ and $\boldsymbol{k}_2$. Results are given} in units $(\alpha Z)^2r_0$, where $r_0 = 2.818$ fm is the classical electron radius.} \label{RealAmpC}
\end{center}
\end{figure}
These spurious contributions should disappear if one would perform the summation over all multipole components of the electron propagator, characterized by the Dirac quantum number $\kappa$. For example, the integral along the path $C_2$ should not contribute to the real part of $M^{(D)}$ at all after summing over all multipoles. This is not the case, however, for the truncated summation over $\kappa$, see Ref.~\cite{scherdin_coulomb_1995}. In order to solve the problem of the spurious contributions, we add the integrals over the segments $C_1$, $C_2$ and $C_3$ for each Dirac quantum number and only after this perform the summation over $\kappa$. This results in a significant acceleration of the convergence of the partial wave summation since the unphysical contributions cancel between the different segments of the integration contour, see Fig.~\ref{RealAmpC}. As seen from this figure, the contributions to the amplitude Re$[\widetilde{M}^{(D)}_\parallel]$ that are obtained upon integration over individual segments $C_1$ (orange dash-dotted line), $C_2$ (blue dashed line) and $C_3$ (red dotted line) exhibit a strong oscillatory behaviour as functions of the scattering angle $\theta$, which usually indicates convergence problems. When summed together (black solid line), however, they provide a reliable prediction for the scattering amplitude which monotonically decreases with $\theta$ and which remains unaltered when including more multipoles. By using this approach, we are able to achieve a relative uncertainty \sfm{for Re$[\widetilde{M}^{(D)}]$ of less than 2\% for $\theta \leq 120^\circ$ and around 5\% for $\theta > 120^\circ$ as well as a relative uncertainty of less than 1\% for Im$[\widetilde{M}^{(D)}]$} by summing over $\kappa$ in the range from $-15$ to $+15$. \sfm{The convergence of the computational results with increasing range of the electron partial waves is illustrated in Fig.~\ref{ConvK}.}

\begin{figure}
\begin{center}
\includegraphics[width=1.\textwidth]{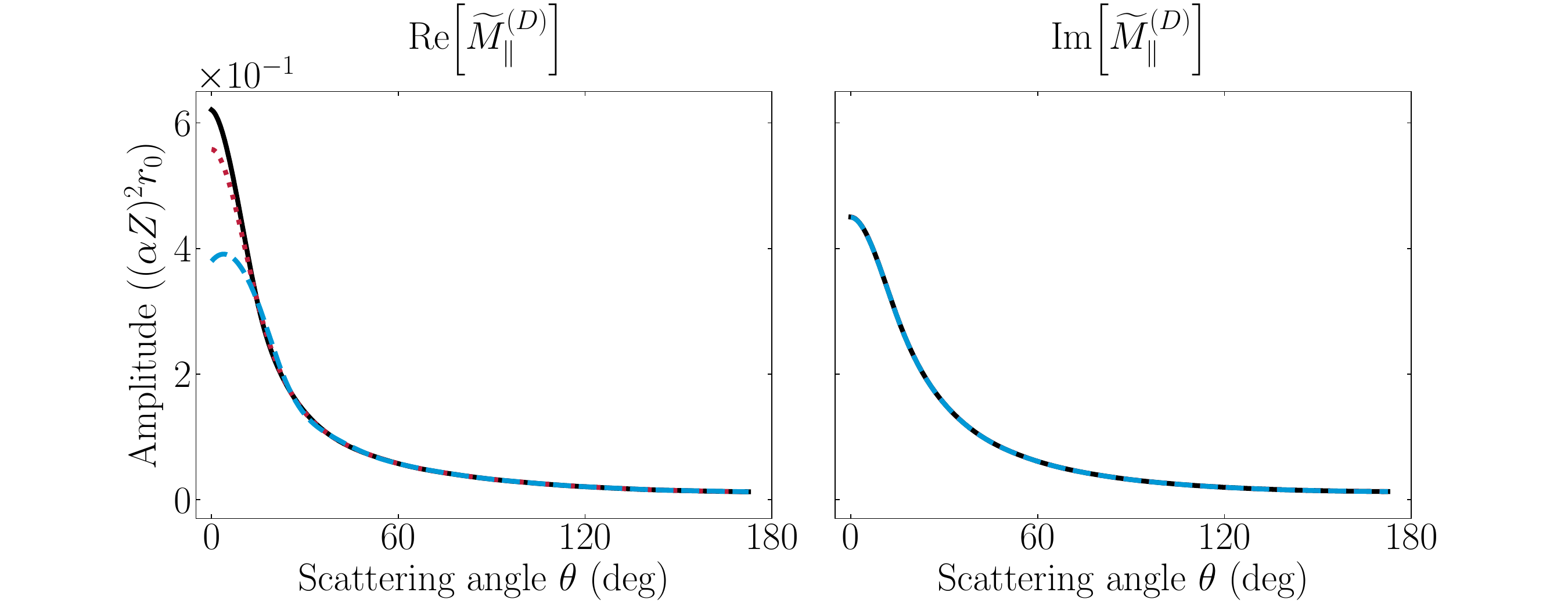} 
\caption{Real (left panel) and imaginary (right panel) part of the Delbrück amplitude~\eqref{delbrueck_amplitude_finite} for the scattering of 2.754~MeV photons by bare plutonium nuclei. The calculations were done for radiation that is linearly polarized within the scattering plane and by summing over the residue of the poles in Eq.~\eqref{intReplace} up to $n = 7$ (blue dashed line), $n = 13$ (red dotted line) and $n = 19$ (black solid line). Results are given in units $(\alpha Z)^2r_0$, where $r_0 = 2.818$ fm is the classical electron radius.} \label{ConvN}
\end{center}
\end{figure}

Another problem with the above-threshold calculations of Delbrück scattering arises due to the poles $C_3$ that are enclosed by the integration contour. These poles are accounted for by summing over the residue in Eq.~\eqref{intReplace}. This summation over the principal quantum number $n$ is infinite reflecting the bound-state spectrum of the hydrogen-like system. The sum converges however relatively fast and, in analogy to Rayleigh scattering~\cite{PhysRevA.107.012805}, high-$n$ terms contribute only for very low scattering angles, see Fig.~\ref{ConvN}. For all calculations presented in the next section, the summation up to $n = 13$ is performed to achieve a relative accuracy of less than 0.1\% for scattering angles larger than $\theta = 30^\circ$.

\sfm{The calculation of the radial integrals~\eqref{IntI} both, near the origin $0 \leq r_1 \leq R_1$, $0 \leq r_2 \leq R_2$, and in the asymptotic regime, relies on the stable evaluation of the involved special functions. In the present work, we employ the \texttt{Arb} C library as implemented by Johansson~\cite{johansson_arb:_2017}, which is based on using the power series representation of these functions. For example, the confluent hypergeometric function, which is related to the Whittaker functions in Eq.~\eqref{IntI}, can be represented as
\begin{equation} \label{conf_hyp}
_1F_1(a;b;z) = \sum_{n=0}^\infty \frac{(a)_nz^n}{(b)_nn!}~.
\end{equation}
The summation in Eq.~\eqref{conf_hyp} runs up to infinity and suffers from severe cancellation problems. Similar problems arise for the sum over the weighted incomplete gamma functions shown in Appendix A. Furthermore, the subtraction of the free-loop contribution in Eq.~\eqref{delbrueck_amplitude_finite} can also cancel many additional digits of accuracy. Therefore, performing computations using the standard double precision arithmetics would lead to the loss of all significant digits in the final Delbrück amplitudes. To overcome this problem, we use arbitrary precision ball arithmetics as implemented in the \texttt{Arb} C library, which is a form of interval arithmetic and, hence, automatically gives rigorous bounds for all rounding errors~\cite{johansson_arb:_2017}. Computations are simplified by using symmetry properties to transform the functions to numerically more favourable parameter regimes. For example, for the confluent hypergeometric function, Kummer's transformation
\begin{equation}
_1F_1(a,b;z) = e^z {_1F_1}(b-a,b;-z)~,
\end{equation}
is employed to avoid calculations with negative arguments which are more susceptible to cancellations errors. We usually use around 25 bytes of precision for the real and complex part of all variables in our calculations and re-evaluate the special functions with a higher precision if the rounding errors get too large.}

Despite all analytical and numerical methods introduced in this and the previous sections, the calculation of Delbrück amplitudes for photon energies above the pair creation threshold requires dramatically more computer time compared to the below threshold case. This is due to the necessity to calculate the radial integrals~\eqref{IntI} close to the edges of the branch cuts at $z' - \tfrac{\omega}{2} = \pm 1$ and $z' + \tfrac{\omega}{2} = \pm 1$, where the integrand is a fast oscillating function that converges particularly slow. To speed up the calculations, we use a hybrid parallelization scheme and calculate the radial integrals with multiple threads using the C library \texttt{pthread}. Moreover, we use \texttt{MPI} to distribute the radial integrals for different energies $z'$ over different nodes of the PTB high perfomance cluster~\cite{mpi40}. Using this method, we can calculate one full set of amplitudes including all scattering angles and polarization states in about one week using around 200 threads.

\section{Numerical Results} \label{NumRes}
\begin{figure}
\begin{center}
\includegraphics[width=1.\textwidth]{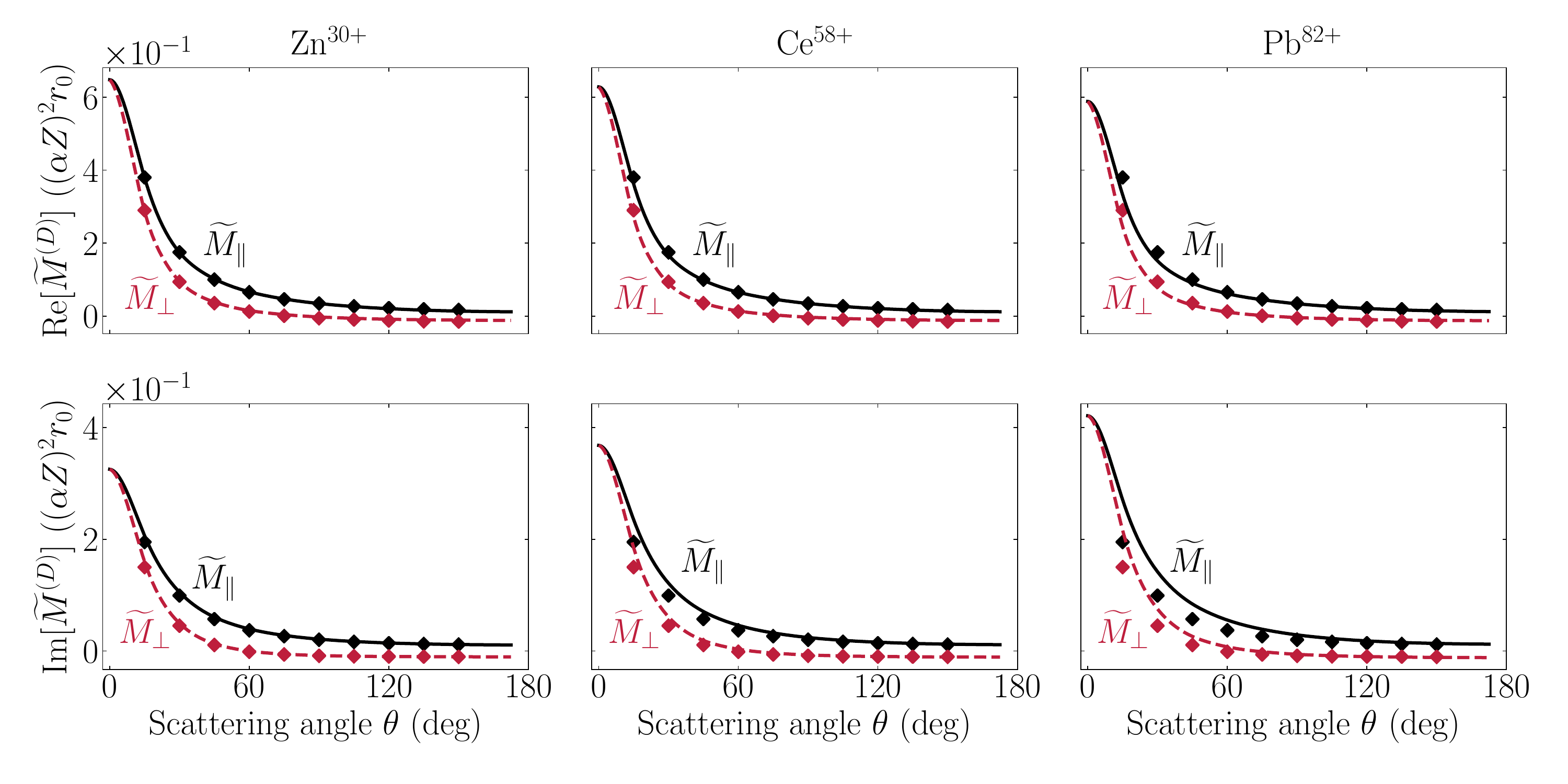} \caption{Real (upper panels) and imaginary (lower panels) parts of the Delbrück amplitude~\eqref{delbrueck_amplitude_finite} for the scattering of 2.754~MeV photons by bare Zn$^{30+}$ (left panels), Ce$^{58+}$ (middle panels) and Pb$^{82+}$ (right panels) ions. For each scenario, calculations have been performed for incoming/outgoing photons that are linearly polarized within (black solid line) or perpendicular (red dashed line) to the scattering plane. Results are given in units $(\alpha Z)^2r_0$, where $r_0 = 2.818$ fm is the classical electron radius.} \label{AllAmp}
\end{center}
\end{figure}

Above we have discussed the theoretical and computational details of calculating Delbrück scattering amplitudes for photon energies above the $e^+e^-$ pair production threshold. To illustrate the application of our method, we will present numerical results for the scattering of 2.754~MeV photons off bare zinc, cerium and lead ions. The interest to this energy and nuclear charge range arises from a series of experiments in which gamma rays, emitted from radioactive sources, were elastically scattered by atomic targets~\cite{SCHUMACHER1999101, SCHUMACHER1975134,RULLHUSEN1979166,rullhusen_coulomb_1979, PhysRevC.23.1375}. In these experiments, the scattering cross section was measured for a wide range of emission angles $\theta$ of the final-state photon. To investigate such an angle-differential cross section, one needs to know the real and imaginary parts of the Delbrück amplitude as a function of $\theta$ and for different polarizations of the scattered photons. In Fig.~\ref{AllAmp}, we present $\widetilde{M}^{(D)}_\parallel(\theta)$ (black solid line) and $\widetilde{M}^{(D)}_\perp(\theta)$ (red dashed line), which describe the cases where the incoming and outgoing photons are linearly polarized either within or perpendicular to the scattering plane spanned by wave vectors $\boldsymbol{k}_1$ and $\boldsymbol{k}_2$. As known from symmetry considerations, these two independent amplitudes are sufficient to predict all observables for the scattering of light off a spherically symmetric system~\cite{PhysRevA.13.692,PhysRevA.34.1178}.  

In Fig.~\ref{AllAmp}, moreover, we compare our numerical results with the predictions of the lowest-order Born approximation (diamonds), which is described by the second Feynman diagram on the right-hand side of Fig.~\ref{delbrueck_expansion} and is of the order $(\alpha Z)^2$. By comparing all-order and Born calculations, we can investigate the role of the Coulomb corrections, the first of which is displayed by the Feynman diagram in the second line of Fig.~\ref{delbrueck_expansion}. 
As seen from the left column of Fig.~\ref{AllAmp}, the results for the all-order calculations match the predictions by the lowest-order Born approximation in the low-$Z$ regime, i.e. for the Zn$^{30+}$ ion. This is well expected as the higher-order Coulomb corrections are negligible if $\alpha Z$ is small. These corrections, however, rapidly grow with increase of the nuclear charge as $(\alpha Z)^4$ in the leading order. As seen from the middle and right panels, the Coulomb corrections become visible for medium- and high-$Z$ ions leading to a slight reduction of the real part of the Delbrück amplitude and a strong enhancement of its imaginary part. For $\theta =45^\circ$, for example, Im$[\widetilde{M}^{(D)}_\perp]$ is enhanced by a factor of $1.8$ and $2.5$ for the scattering off bare cerium and lead ions, respectively.

\begin{figure}
\begin{center}
\includegraphics[width=1.\textwidth]{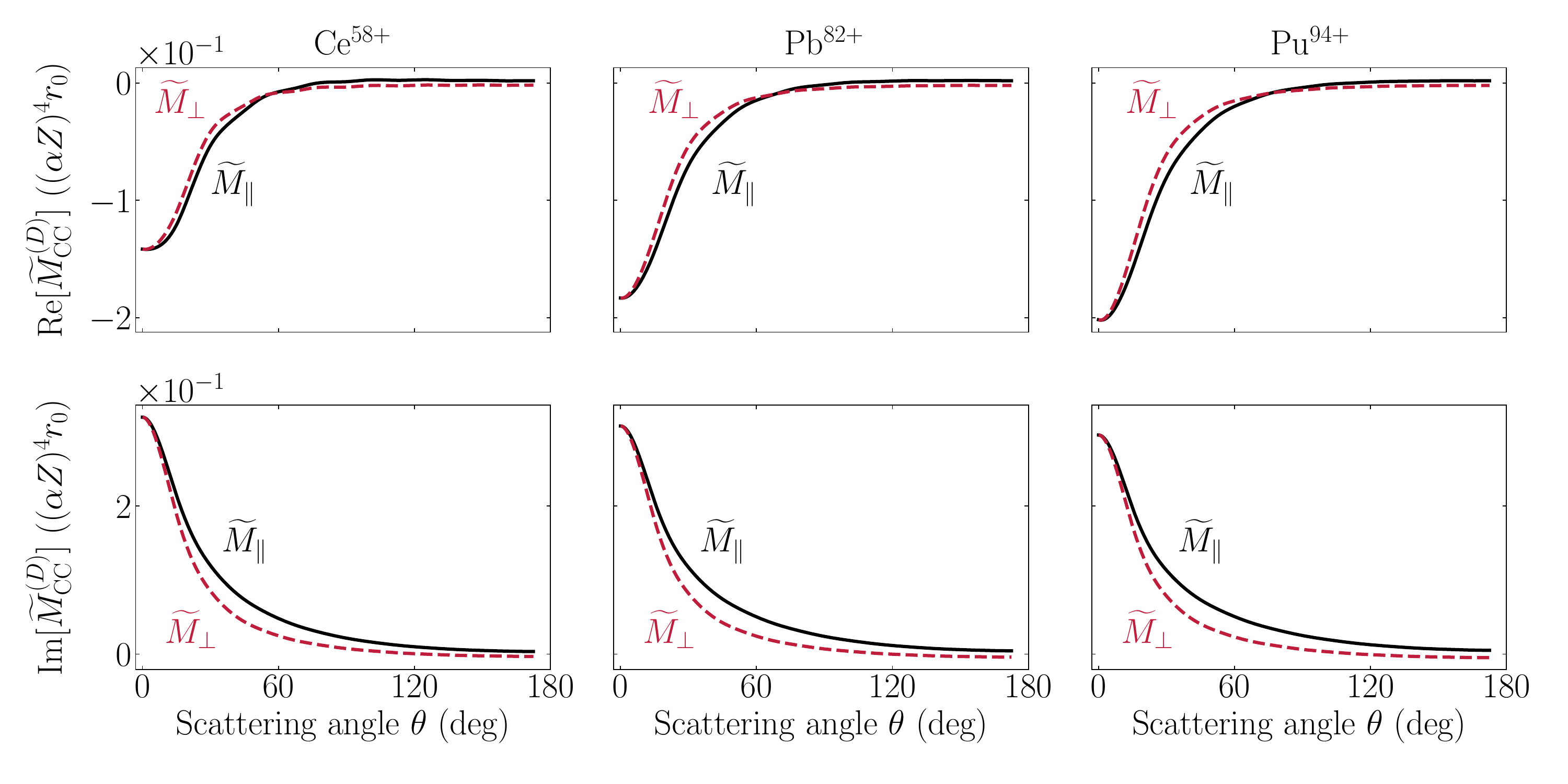} \caption{Real (upper panels) and imaginary (lower panels) parts of the Coulomb corrections to the Delbrück amplitude~\eqref{delbrueck_amplitude_finite} for the scattering of 2.754~MeV photons by bare Ce$^{58+}$ (left panels), Pb$^{82+}$ (middle panels) and Pu$^{94+}$ (right panels) ions. For each scenario, calculations have been performed for incoming/outgoing photons that are linearly polarized within (black solid line) or perpendicular (red dashed line) to the scattering plane. Results are given in units $(\alpha Z)^4r_0$, where $r_0 = 2.818$ fm is the classical electron radius.} \label{AmpCC}
\end{center}
\end{figure}

To better understand the nuclear charge and angular behaviour of the Coulomb corrections, we display in Fig.~\ref{AmpCC} the difference between the all-order and the lowest-order Born calculations for the scattering off bare cerium, lead and plutonium ions. The results for zinc are not presented here since the Coulomb corrections are very small for the low-$Z$ regime and, for this reason, their accurate evaluation is very cumbersome and requires a huge computation time. As seen from the figure, the higher-order corrections roughly scale as $(\alpha Z)^4$ which is well expected since the first term beyond the lowest-order Born approximation obeys this scaling behaviour, see Fig.~\ref{delbrueck_expansion}. However, one can also observe a slight difference between the scaled Coulomb corrections for various elements that originates from even higher-order contributions of the order $(\alpha Z)^6$ and beyond. As seen from Fig.~\ref{AmpCC}, these contributions lead to a remarkable enhancement of the absolute value of the real part of the Coulomb corrections but only slightly affects the imaginary part of $\widetilde{M}^{(D)}$.

\section{Conclusion} \label{Conc}
In conclusion, we have presented a theoretical method for accurate and numerically stable calculations of Delbrück scattering amplitudes for photon energies above the electron-positron pair production threshold. This method takes into account the interaction of the virtual $e^+e^-$ pair with the Coulomb field of a nucleus to all orders in the interaction strength parameter $\alpha Z$. In order to perform such an all-order analysis, we made use of the relativistic Coulomb Green's function approach. The evaluation of the resulting amplitude requires \sfm{performing numerical integrations over the energy and radial arguments} of the Green functions which \sfm{is a rather demanding task}. We carry out the integration over the loop energy by using a modified Wick rotated integration contour and improve the numerical stability of our approach by solving the integrals over the radial vertex coordinates analytically in the asymptotic regime.

For the illustration of the use of the proposed method, detailed calculations for the scattering of $2.754$~MeV photons off bare zinc, cerium, lead and plutonium ions were performed. In these calculations, we paid special attention to the role of the Coulomb corrections to the scattering amplitude beyond the first-order Born approximation. The Coulomb corrections, whose leading order is $(\alpha Z)^4$, were found to affect mainly the imaginary part of the Delbrück amplitude. In particular, Im$[\widetilde{M}^{(D)}]$ can be enhanced by almost a factor of three in the high-$Z$ regime. Our calculations clearly demonstrate that the accurate treatment of the Coulomb corrections to Delbrück scattering are strongly demanded for the guidance and planning of future gamma-ray scattering experiments as planned using synchrotron facilities and radiative sources.

\section*{Acknowledgments}
This work has been supported by the GSI Helmholtz Centre for Heavy Ion Research under the project BSSURZ1922. We thank our colleagues from PTB's high performance computing division and especially Gert Lindner for providing access to their computation cluster and for their excellent technical support. We also thank Sebastian Ulbricht, Sophia Strnat and Olli for very helpful discussions.

\appendix

\section{Analytical solution of the radial integrals} \label{AnaInt}

In analogy to Ref.~\cite{PhysRevA.105.022804}, we split the integral in Eq.~\eqref{IntI} into parts that are close and far away from the origin and insert the asymptotic expansion of the Whittaker functions in the outer part. The expression that needs to be solved analytically reads

\begin{equation}
\begin{aligned}
&\int_{R_1}^\infty \frac{\text{d}r_1}{r_1} W_{\nu'+\frac{p_1}{2},\lambda'}(2c'r_1) j_{L_1}(\omega r_1) W_{\nu+\frac{p_2}{2},\lambda}(2cr_1)\\
&\times\Big[\mathcal{C}\! +\!\!\int_{R_2}^{r_1} \frac{\text{d}r_2}{r_2} M_{\nu'+\frac{p_3}{2},\lambda'}(2c'r_2) j_{L_2}(\omega r_2) M_{\nu+\frac{p_4}{2},\lambda}(2cr_2)\Big]\\
&\to (2c')^{\nu'+p_1/2}(2c)^{\nu+p_2/2}\left\{\text{I}\times\mathcal{C} + \sum_{j=1}^{4}  \alpha_j\left[\text{III}_j- \text{I}\times\text{II}_j\right] \right\}~,
\end{aligned}
\end{equation}

\noindent where

\begin{equation}
\begin{aligned}
\alpha_1 &= \frac{\Gamma(1+2\lambda)\Gamma(1+2\lambda')(2c')^{-\nu'-p_3/2}(2c)^{-\nu-p_4/2}}{\Gamma(\frac{1}{2}+\lambda-\nu-\frac{p_4}{2})\Gamma(\frac{1}{2}+\lambda'-\nu'-\frac{p_3}{2})}\\
\alpha_2 &= \frac{\Gamma(1+2\lambda)\Gamma(1+2\lambda')(2c')^{\nu'+p_3/2}(2c)^{\nu+p_4/2}}{\Gamma(\frac{1}{2}+\lambda+\nu+\frac{p_4}{2})\Gamma(\frac{1}{2}+\lambda'+\nu'+\frac{p_3}{2})}e^{q'(\frac{1}{2}+\lambda'-\nu'-\frac{p_3}{2})\pi i + q(\frac{1}{2}+\lambda-\nu-\frac{p_4}{2})\pi i}\\
\alpha_3 &= \frac{\Gamma(1+2\lambda)\Gamma(1+2\lambda')(2c')^{-\nu'-p_3/2}(2c)^{\nu+p_4/2}}{\Gamma(\frac{1}{2}+\lambda+\nu+\frac{p_4}{2})\Gamma(\frac{1}{2}+\lambda'-\nu'-\frac{p_3}{2})}e^{q(\frac{1}{2}+\lambda-\nu-\frac{p_4}{2})\pi i}\\
\alpha_4 &= \frac{\Gamma(1+2\lambda)\Gamma(1+2\lambda')(2c')^{\nu'+p_3/2}(2c)^{-\nu-p_4/2}}{\Gamma(\frac{1}{2}+\lambda'+\nu'+\frac{p_3}{2})\Gamma(\frac{1}{2}+\lambda-\nu-\frac{p_4}{2})}e^{q'(\frac{1}{2}+\lambda'-\nu'-\frac{p_3}{2})\pi i}
\end{aligned}
\end{equation}

\begin{equation} \label{IntToSolve}
\begin{aligned}
\text{I} = &\sum_{s_W', s_W = 0}^\infty (-2c')^{-s_W'} u_W(s_W', \nu'+\tfrac{p_1}{2}, \lambda')(-2c)^{-s_W}u_W(s_W, \nu+\tfrac{p_2}{2}, \lambda)\\
&\times\int_{R_1}^\infty \text{d}r_1~e^{-(c+c')r_1} r_1^{-1+\nu'+\nu+(p_1+p_2)/2-s_W'-s_W}j_{L_1}(\omega r_1)~,\\
\text{II}_j = &\sum_{s_M', s_M = 0}^\infty \beta'_j\beta_j\Bigg[\left.\int \text{d}r_2~e^{(\gamma_jc+\gamma'_jc')r_2} r_2^{-1-\gamma'_j\nu'-\gamma_j\nu-(\gamma'_jp_3+\gamma_jp_4)/2-s_M'-s_M}j_{L_2}(\omega r_2)\Bigg]\right\vert_{r_2=R_2}~,\\
\text{III}_j = &\sum_{s_W', s_W, s_M', s_M = 0}^\infty(-2c')^{-s_W'} u_W(s_W', \nu'+\tfrac{p_1}{2}, \lambda')(-2c)^{-s_W}u_W(s_W, \nu+\tfrac{p_2}{2}, \lambda)\beta'_j\beta_j \\
&\times\int_{R_1}^\infty \text{d}r_1~e^{-(c+c')r_1} r_1^{-1+\nu'+\nu+(p_1+p_2)/2-s_W'-s_W}j_{L_1}(\omega r_1)\\
&\times \Bigg[\left.\int \text{d}r_2~e^{(\gamma_jc+\gamma'_jc')r_2} r_2^{-1-\gamma'_j\nu'-\gamma_j\nu-(\gamma'_jp_3+\gamma_jp_4)/2-s_M'-s_M}j_{L_2}(\omega r_2)\Bigg]\right\vert_{r_2=r_1}~,
\end{aligned}
\end{equation}

\begin{equation}
\begin{aligned}
\beta'_1 &= (2c')^{-s_M'}u_M(s_M',\nu'+\tfrac{p_3}{2},\lambda'),~\beta_1 = (2c)^{-s_M}u_M(s_M,\nu+\tfrac{p_4}{2},\lambda)~,\\
\beta'_2 &= (-2c')^{-s_M'}\widetilde{u}_M(s_M',\nu'+\tfrac{p_3}{2},\lambda'),~\beta_2 = (-2c)^{-s_M}\widetilde{u}_M(s_M,\nu+\tfrac{p_4}{2},\lambda)~,\\
\beta'_3 &= (2c')^{-s_M'}u_M(s_M',\nu'+\tfrac{p_3}{2},\lambda'),~\beta_3 = (-2c)^{-s_M}\widetilde{u}_M(s_M,\nu+\tfrac{p_4}{2},\lambda)~,\\
\beta'_4 &= (-2c')^{-s_M'}\widetilde{u}_M(s_M',\nu'+\tfrac{p_3}{2},\lambda'),~\beta_4 = (2c)^{-s_M}u_M(s_M,\nu+\tfrac{p_4}{2},\lambda)~,
\end{aligned}
\end{equation}

\begin{equation}
\begin{aligned}
\gamma'_1 &= +1,~\gamma_1 &= +1~,\\
\gamma'_2 &= -1,~\gamma_2 &= -1~,\\
\gamma'_3 &= +1,~\gamma_3 &= -1~,\\
\gamma'_4 &= -1,~\gamma_4 &= +1~,\\
\end{aligned}
\end{equation}

\noindent and

\begin{equation}
\mathcal{C} = \int_0^{R_2} \frac{\text{d}r_2}{r_2} M_{\nu'+\frac{p_3}{2},\lambda'}(2c'r_2) j_{L_2}(\omega r_2) M_{\nu+\frac{p_4}{2},\lambda}(2cr_2)~.
\end{equation}

Solving the integrals in Eq.~\eqref{IntToSolve} is completely analogous to the case where just the first term of the asymptotic expansion of the Whittaker function contributes. Following the same steps as in Ref.~\cite{PhysRevA.105.022804}, we obtain for the integral that occurs in $\text{III}_j$

\begin{equation}
\begin{aligned}
\int_{R_1}^\infty& \text{d}r_1~e^{-(c+c')r_1} r_1^{-1+\nu'+\nu+(p_1+p_2)/2-s_W'-s_W}j_{L_1}(\omega r_1)\\
\times& \Bigg[\left.\int \text{d}r_2~e^{(\gamma_jc+\gamma'_jc')r_2} r_2^{-1-\gamma'_j\nu'-\gamma_j\nu-(\gamma'_jp_3+\gamma_jp_4)/2-s_M'-s_M}j_{L_2}(\omega r_2)\Bigg]\right\vert_{r_2=r_1}\\
=& \sum_{m_1=0}^{L_1} \sum_{m_2=0}^{L_2} \frac{(L_1+m_1)!}{m_1!(L_1-m_1)!} \frac{(L_2+m_2)!}{m_2!(L_2-m_2)!} i^{L_1+L_2+2-m_1-m_2} (2\omega)^{-m_1-m_2-2}\\
&\times [(-1)^{L_1+L_2+2-m_1-m_2}N_{j,++}+(-1)^{L_1+1-m_1}N_{j,+-}+(-1)^{L_2+1-m_2}N_{j,-+}+N_{j,--}]~,
\end{aligned}
\end{equation}

\noindent where

\begin{equation}
\begin{aligned}
N_{j,++} &= \sum_{s_G=0}^\infty\frac{(2+\gamma'_j\nu'+\gamma_j\nu+(\gamma'_jp_3+\gamma_jp_4)/2+m_2+s_M'+s_M)_{s_G}}{(c-\gamma_jc+c'-\gamma'_jc'-2i\omega)(\gamma_jc+\gamma'_jc'+ i\omega)^{s_G+1}}\\
&\times (c\!-\!\gamma_jc\!+\!c'\!\!-\!\gamma'_jc'\!\!-\!2i\omega)^{4-(p_1+p_2-\gamma'_jp_3-\gamma_jp_4)/2-\nu+\gamma_j\nu-\nu'\!+\gamma'_j\nu'\!+m_1+m_2+s_M'\!+s_M+s_W'+s_W+s_G}\\
&\times\Gamma(-3+(p_1+p_2-\gamma'_jp_3+\gamma_jp_4)/2+\nu-\gamma_j\nu+\nu'-\gamma'_j\nu'\\
&\hphantom{\times\Gamma(}-m_1-m_2-s_M'-s_M-s_W'-s_W-s_G, (c-\gamma_jc+c'-\gamma'_jc'-2i\omega) R_1)~,\\
N_{j,--} &= \sum_{s_G=0}^\infty\frac{(2+\gamma'_j\nu'+\gamma_j\nu+(\gamma'_jp_3+\gamma_jp_4)/2+m_2+s_M'+s_M)_{s_G}}{(c-\gamma_jc+c'-\gamma'_jc'+2i\omega)(\gamma_jc+\gamma'_jc'- i\omega)^{s_G+1}}\\
&\times (c\!-\!\gamma_jc\!+\!c'\!\!-\!\gamma'_jc'\!\!+\!2i\omega)^{4-(p_1+p_2-\gamma'_jp_3-\gamma_jp_4)/2-\nu+\gamma_j\nu-\nu'\!+\gamma'_j\nu'\!+m_1+m_2+s_M'\!+s_M+s_W'+s_W+s_G}\\
&\times\Gamma(-3+(p_1+p_2-\gamma'_jp_3-\gamma_jp_4)/2+\nu-\gamma_j\nu+\nu'-\gamma'_j\nu'\\
&\hphantom{\times\Gamma(}-m_1-m_2-s_M'-s_M-s_W'-s_W-s_G, (c-\gamma_jc+c'-\gamma'_jc'+2i\omega) R_1)~,\\
N_{j,+-} &= \sum_{s_G=0}^\infty\frac{(2+\gamma'_j\nu'+\gamma_j\nu+(\gamma'_jp_3+\gamma_jp_4)/2+m_2+s_M'+s_M)_{s_G}}{(c-\gamma_jc+c'-\gamma'_jc')(\gamma_jc+\gamma'_jc'- i\omega)^{s_G+1}}\\
&\times (c\!-\!\gamma_jc\!+\!c'\!\!-\!\gamma'_jc')^{4-(p_1+p_2-\gamma'_jp_3-\gamma_jp_4)/2-\nu+\gamma_j\nu-\nu'+\gamma'_j\nu'+m_1+m_2+s_M'+s_M+s_W'+s_W+s_G}\\
&\times\Gamma(-3+(p_1+p_2-\gamma'_jp_3-\gamma_jp_4)/2+\nu-\gamma_j\nu+\nu'-\gamma'_j\nu'\\
&\hphantom{\times\Gamma(}-m_1-m_2-s_M'-s_M-s_W'-s_W-s_G, (c-\gamma_jc+c'-\gamma'_jc') R_1)~,\\
N_{j,-+} &= \sum_{s_G=0}^\infty\frac{(2+\gamma'_j\nu'+\gamma_j\nu+(\gamma'_jp_3+\gamma_jp_4)/2+m_2+s_M'+s_M)_{s_G}}{(c-\gamma_jc+c'-\gamma'_jc')(\gamma_jc+\gamma'_jc'+ i\omega)^{s_G+1}}\\
&\times (c\!-\!\gamma_jc\!+\!c'\!\!-\!\gamma'_jc')^{4-(p_1+p_2-\gamma'_jp_3-\gamma_jp_4)/2-\nu+\gamma_j\nu-\nu'+\gamma'_j\nu'+m_1+m_2+s_M'+s_M+s_W'+s_W+s_G}\\
&\times\Gamma(-3+(p_1+p_2-\gamma'_jp_3-\gamma_jp_4)/2+\nu-\gamma_j\nu+\nu'-\gamma'_j\nu'\\
&\hphantom{\times\Gamma(}-m_1-m_2-s_M'-s_M-s_W'-s_W-s_G, (c-\gamma_jc+c'-\gamma'_jc') R_1)~.\\
\end{aligned}
\end{equation}

\noindent The case $j = 1$ has to be handled separately for $N_{j,-+}$ and $N_{j,-+}$. However, exactly these integrals were already calculated in Ref.~\cite{PhysRevA.105.022804}. The integrals that occur in $\text{I}$ and $\text{II}_j$ are given by

\begin{equation} \label{Intr2}
\begin{aligned}
\int_{R_1}^\infty&\text{d}r_1~e^{-(c+c')r_1} r_1^{-1+\nu'+\nu+(p_1+p_2)/2-s_W'-s_W}j_{L_1}(\omega r_1) \\
&= \sum_{m_1=0}^{L_1} \frac{(L_1+m_1)!}{m_1!(L_1-m_1)!} i^{L_1+1-m_1} (2\omega)^{-m_1-1}[(-1)^{L_1+1-m_1}O_++O_-]~,
\end{aligned}
\end{equation}

\noindent where

\begin{equation}
\begin{aligned}
O_{\pm} &= \frac{(c+c'\mp i\omega)^{2-\nu-\nu'-(p_1+p_2)/2+m_1+s_W'+s_W}}{c+c'\mp i\omega}\\
&\times \Gamma(-1+\nu'+\nu+(p_1+p_2)/2-m_1-s_W'-s_W, +(c+c'\mp i\omega)R_1)~,\\
\end{aligned}
\end{equation}

\noindent and

\begin{equation} \label{r1Int}
\begin{aligned}
\int&\text{d}r_2~e^{+(\gamma_jc+\gamma'_jc')r_2} \left.r_2^{-1-\gamma'_j\nu'-\gamma_j\nu-(\gamma'_jp_3+\gamma_jp_4)/2-s_M'-s_M}j_{L_2}(\omega r_2)\right\vert_{r_2=R_2} \\
&= \sum_{m_2=0}^{L_2} \frac{(L_2+m_2)!}{m_2!(L_2-m_2)!} i^{L_2+1-m_2} (2\omega)^{-m_2-1}[(-1)^{L_2+1-m_2}N_++N_-]~,
\end{aligned}
\end{equation}

\noindent where

\begin{equation}
\begin{aligned}
N_{\pm} &= \frac{(-\gamma_jc-\gamma'_jc'\mp i\omega)^{2+\gamma_j\nu+\gamma'_j\nu'+(\gamma'_jp_3+\gamma_jp_4)/2+m_2+s_M'+s_M}}{\gamma_jc+\gamma'_jc'\pm i\omega}\\
&\times \Gamma(-1-\gamma'_j\nu'-\gamma_j\nu-(\gamma'_jp_3+\gamma_jp_4)/2-m_2-s_M'-s_M, -(\gamma_jc+\gamma'_jc'\pm i\omega)R_2)~,\\
\end{aligned}
\end{equation}

\noindent respectively.

\bibliographystyle{apsrev4-1}

\end{document}